\documentclass{article}
\usepackage{amssymb,amsmath,cite,bm}

\usepackage{graphicx}

\begin{document}

\noindent{\LARGE\bfseries Permutation approach, high frequency trading and variety of micro patterns in financial time series}\\[2\baselineskip]
\textsf{Cina~Aghamohammadi}$^{a}$~\footnote{e-mail: caghamohammadi@ucdavis.edu },
\textsf{Mehran~Ebrahimian}$^{b}$~\footnote{e-mail: m\_ebrahimian@gsme.sharif.edu },
\textsf{Hamed~Tahmooresi}$^{c}$~\footnote{e-mail: tahmooresi@ce.sharif.edu},
\\[2\baselineskip]
$^a$Physics Department, University of California, Davis, California 95616, USA\\
$^b$Graduate school of management and economic, Sharif University of Technology, 8639-11155, Tehran, Iran\\
$^c$Department of Computer Engineering, Sharif University of Technology, 11155-9517, Tehran, Iran


\begin
{abstract}
Permutation approach is suggested as a method to investigate financial time series in 
micro scales. The  method is used to see how high frequency trading in recent years has affected the micro patterns which may be seen in financial time series.
Tick to tick exchange rates are considered as examples.
It is seen that variety of patterns evolve through time; and 
that the scale over which the target markets have no dominant patterns, have decreased steadily over time with the emergence of higher frequency trading.

\end{abstract}
\newpage
\section{Introduction}
In the middle of sixties,  the algorithmic
complexity theory was independently developed by  Kolmogorov \cite{KOl} and Chaitin \cite{Cha}. To parameterize complexity in deterministic or random dynamical systems, the most important quantity which may
be used is entropy. There are different ways to count the diversity of any pattern
generated by a data source: Shannon entropy, metric entropy, topological entropy, etc.
After the seminal works of Shannon \cite{SHA}, in 1949 the word entropy came to the fore in the new context of information theory, coding theory, and cryptography. Recently the concept of entropy is also used in econophysics (see \cite{JV,AGol} and references therein) and sociodynamics \cite{DHel}.
The concept of entropy has been evolved along different ways: Renyi
entropy \cite{Ren}, topological entropy \cite{AKM}, Tsallis entropy \cite{CTs}, directional entropy \cite{Mil}, permutation entropy \cite{JA}, epsilon-tau entropy \cite{PRR}, etc.
Permutation entropy was introduced by Bandt, Keller, and Pompe in \cite{BP,BKP}.
Entropies are basic observables for dynamical systems. In \cite{BKP} a piecewise monotone map $f$ from an interval $I$ into itself is defined, and it is shown that for piecewise monotone interval maps the Kolmogorov-Sinai entropy
can be obtained from order statistics of the values in a generic orbit. 
It has been shown that 
it is possible to use the permutation entropy to detect dynamical changes in a complex time series \cite{CTGPH}.

Recently, permutation entropy has been used to study 
dynamical changes of EEG data \cite{LOR}; and
 based on permutation entropy, mutual information of two oscillators 
has been calculated \cite{BGSM}.

In this article variety of micro patterns in financial time series  have been studied. Variety of patterns
 in financial time series is an important measure.
In a completely random series all different patterns may occur with equal weight. If some patterns in a time series are much less than the others, the time series contains dominant patterns. These dominant patterns represent some characteristics of the system, which needs to be revealed.
This information also could be used for prediction of future changes, which for financial time series represent inefficiency in the target market.
There are several definition for the efficient market hypothesis (EMH) \cite{FAM,FAM2,JEN}. According to the EMH hypothesis, 
asset prices move as random walks over time, and technical analysis should provide no useful information to predict future changes \cite{BEE}. This means
that asset prices in an efficient market fluctuate randomly in response to the unanticipated component of news \cite{SAM}. Some people take EMH as a core assumption in finance theory \cite{ROS}. But many physicists consider it only an approximation \cite{FAR}. According to EMH, in 
an efficient market variety of patterns for 
increments of the price in different scales should be at maximum level.

It is now known that most markets behave efficiently in macro scale, 
and there are no dominate patterns in their financial time series. 
But it seems that the investigation of micro patterns, and searching for 
dominant patterns by participants in markets, in recent years,  
have faded those patterns. Permutation entropy is taken as  a criteria for measuring the variety of micro patterns which may be seen in financial time series. Tick to tick exchange rate time series are considered as an example.

\section{Definition}
Consider a set of $n$ distinct real numbers, $\{a_1,a_2,\cdots, a_n\}$. One may define a map from these numbers to the set  $\{1,2,\cdots, n\}$ in such a way that the ordering of the second set is the same as the first one.
The range of this map will be $n!$ permutations. 
The permutation corresponding to $\{a_1, a_2, \dots, a_n\}$ 
is called the \emph{pattern}, and is denoted by $\Pi$. 
See fig. 1, for the case $n=3$.
Consider a time series $\{x_i\}_{i=1,\cdots,N}$. 
By a window of length $n$, we mean any subsequence of the form
$\{a_m, a_{m+1}, \dots, a_{m+n}\}$.
There are $N-n+1$ windows of length $n$, to each of them there 
corresponds a pattern $\Pi$.    
If $\Pi_i$ is a given pattern, we define
\begin{equation}\label{01}
p_i:=\frac{N_i}{N-n+1},
\end{equation}
where $N_i$ is the number of $n$ consecutive numbers with 
pattern $\Pi_i$. 
For large $N$, $p_i$ tends to the probability of occurring the 
pattern $\Pi_i$.
Permutation entropy of order $n$ of a time series, $\{x_k\}_{k=1}^N$, is defined as (see e.g. \cite{BP})
\begin{equation}\label{02}
H_n:=-\sum_{i=1}^{n!}p_i\log p_i.
\end{equation}
It can be shown that $0\leq H_n\leq \log n!$ \cite{CT}. Upper bound occurs when all 
$p_i$'s have the same value, \emph{i.e.} when the time series is a 
completely random series; 
and the lower bound occurs when only one of $p_i$'s is nonzero, which happens
when the time series is a decreasing or increasing sequence.

\begin{figure}
\begin{center}
\includegraphics[scale=0.3]{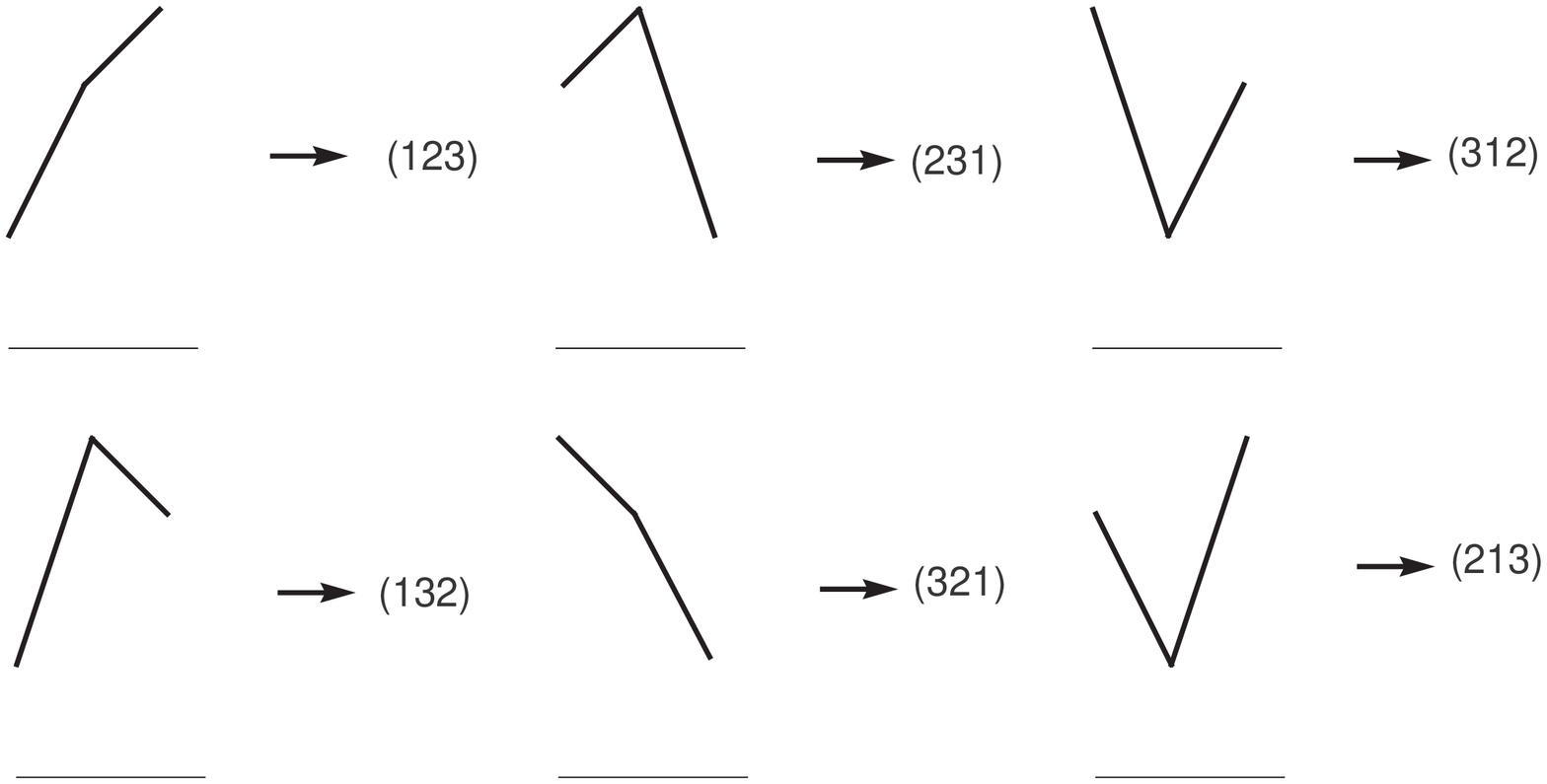}
\setlength{\unitlength}{1pt}
\caption{\label{fig1}}{Permutations $\Pi_i, \ {i=1,2,\cdots, 6}$, for $n=3$ }
\end{center}
\end{figure}

In some cases a linear function
$H_n= k(n-1)+C$ is a good approximation.
This means that for these time series, in the view of permutation entropy, there are 
just two degrees of freedom for the time 
series\footnote{Of course this is not a common case, for example, 
in a completely random time series $H_n
\propto n\, \ln n$, for large $n$, which is not a linear function of $n$.}.

For large $n$, $k$ determines the behavior of  $H_n$. To 
quantize the variety of patterns in this type of time series one may use the parameter $k$ .

It is useful to define a new parameter,
$h_n$ which is the permutations entropy of order $n$ of a time series 
per symbol:
\begin{equation}\label{03}
h_n:=\frac{H_n}{n-1}.
\end{equation}
The denominator is $n-1$, 
because calculating a change means $n>1$.
For large $n$, $h_n$ converges to $k$. Therefore to calculate $k$
one can find the slope of $H_n$ as a function of $n$.

\section{The problem}
If the information embodied in a time series cannot be compressed, the time series is called unpredictable. As it is stated in \cite{RNH}, no difference can be detected between a time series carrying a large amount of non-redundant economic information and a pure random process, and it is well-known that the financial time series carries a large amount of non-redundant information. Therefore, the complexity of time series can be measured by
 assigning a measure to the variety of patterns which are embedded in time series.

\begin{figure}
\begin{center}
\includegraphics[scale=0.6]{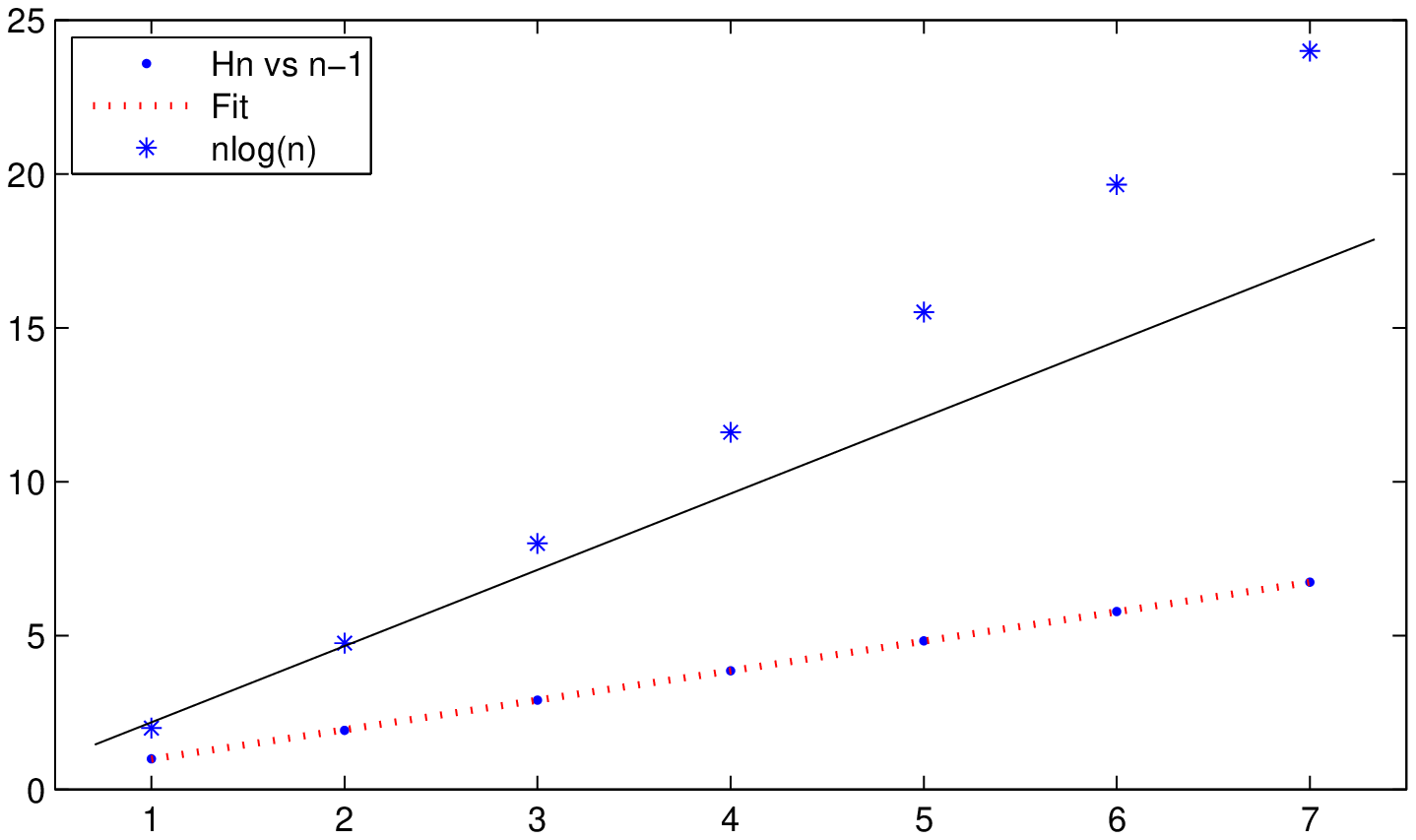}
\setlength{\unitlength}{1pt}
\caption{\label{fig2}}{The plot of $H_n,\ n=2,3,\cdots ,8$, and $n\log(n)$ as a function of $n-1$. In some cases, such as a completely random time series, $H_n $ for large $n$ does not behave as a linear function of $n$, for example in this special case behave as $n\log(n)$. }
\end{center}
\end{figure}

We have analyzed the exchange rates as a time series.
The data we have used, are tick to tick exchange rates of USD/JPY, EUR/GBP and GBP/CHF for 2003 which are time series of about 5000,000 data points, denoted by $x_i$.
Let us calculate $H_n$ for $n=2,\dots, 8$. As it is seen from fig. \ref{fig2}, $H_n$ is a linear function of $n-1$. So it is more useful to work with $h_n$.
To calculate $k$ we need $h_n$ for large $n$. The difference between $h_6$ and $h_7$ is less than 5\%. From now on by large $n$, we mean $n=7$, and we take $h_7$ as an approximation of $k$.

\begin{figure}
\begin{center}
\includegraphics[scale=0.6]{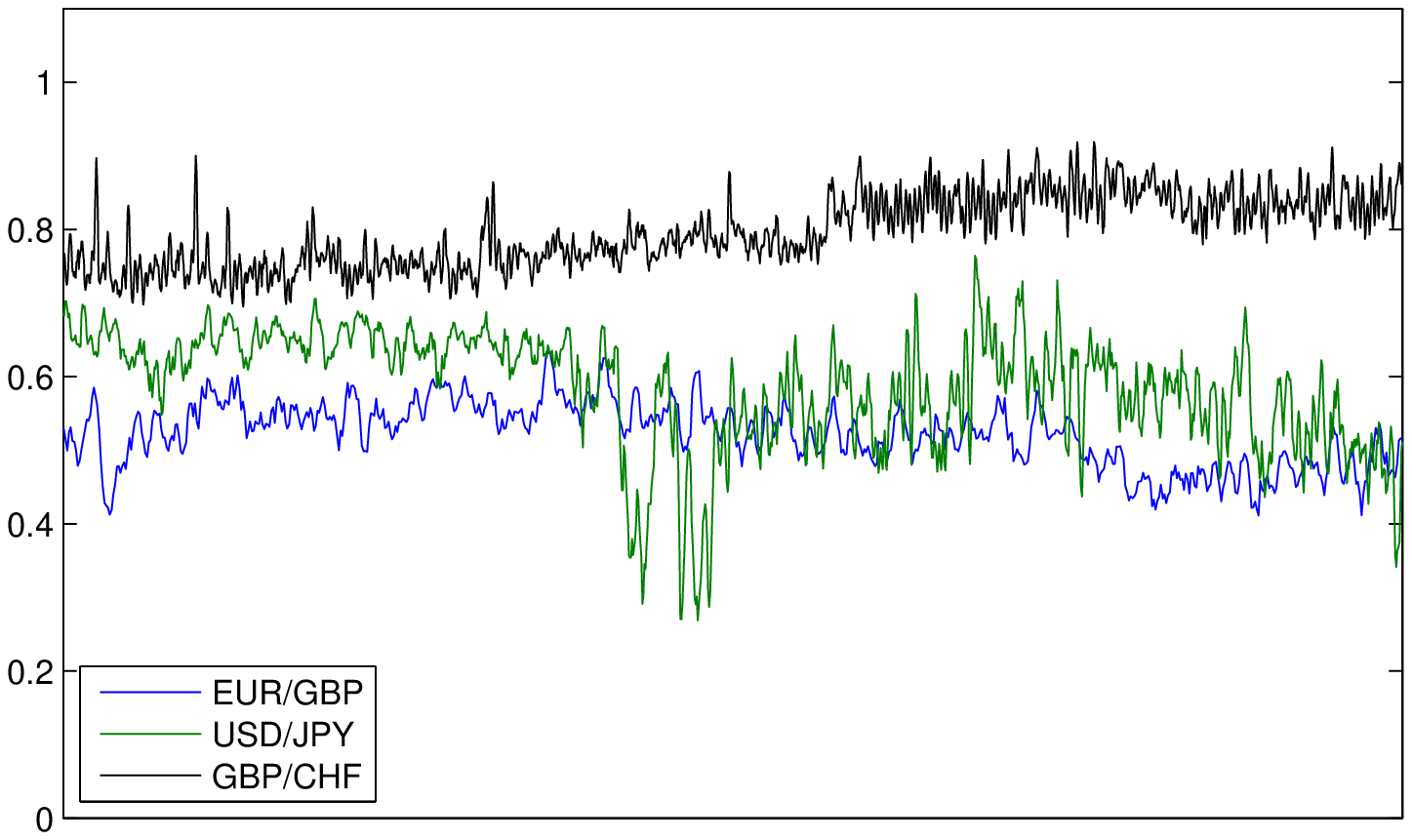} \put(-295,-10){$January\ 2003$}\put(-273,0){$\uparrow$}\put(-56,-10){$january \ 2004$}\put(-34,0){$\uparrow$}\put(-285,147){$\rightarrow$}\put(-380,147){${\rm Pure \ random  \ process}$}
\setlength{\unitlength}{1pt}
\caption{\label{fig4}}{The plot of normalized complexity over time.  $H_7/\log(7!)$ is calculated for windows of length 10000. To obtain the time dependence of
parameter, in each time step the window is shifted by 2000. }
\end{center}
\end{figure}

\section{Variety of micro patterns in exchange rates}

The dynamics of the parameter $k$  may represent the dynamics of complexity of the financial time series over time. It could also be
used as a measure of complexity of financial time series.

One may take a completely random series as a reference series, and compare the complexity of a financial time series with that of
a completely random one.
Since $k$ is not a proper parameter for a completely random 
series, we should take $H_n$ instead of $k$.
To compare the complexity of a time series with
the complexity of completely random series, 
one may normalize it by dividing $H_n / \log(n!)$.
Let us calculate $H_n / \log(n!)$ for windows of length $N=10000$. 
To obtain the time dependence of this parameter, in each time step the window is 
shifted by 2000.
The result for the exchange rate time series is 
shown in fig. \ref{fig4}.
It is seen that tick to tick exchange rates time series show much less variety of micro 
patterns when compared with completely random series.
This means that some patterns occur more than other patterns in these time series. So 
in micro scale some patterns can be seen more than the others.

\section{High frequency trading and its effect on fading patterns}

Recently technology development leads to quick news release, improving the 
quality of data analysis, better information sharing, and increasing the number of 
participants in financial markets. In 2009 high frequency trading 
accounts for over 60\% of the financial exchanges trading volume \cite{ALD}. 
High frequency trading strategies are characterized by a higher 
number of trades and a lower average gain per trade. Currently, there are 
categories of high-frequency trading strategies. Some of them are based on the 
investigation of micro patterns. Typically the holding period of these strategies  are less 
than 1 minute.  The micro patterns observed in
the financial time series have been affected by 
these technology developments.
We believe that the recent investigations 
of the micro patterns, and searching for the
dominant patterns by participants in the markets, 
have faded those patterns.  
We now present our reasoning. 
\par
\begin{figure}
\begin{center}
\includegraphics[scale=0.6]{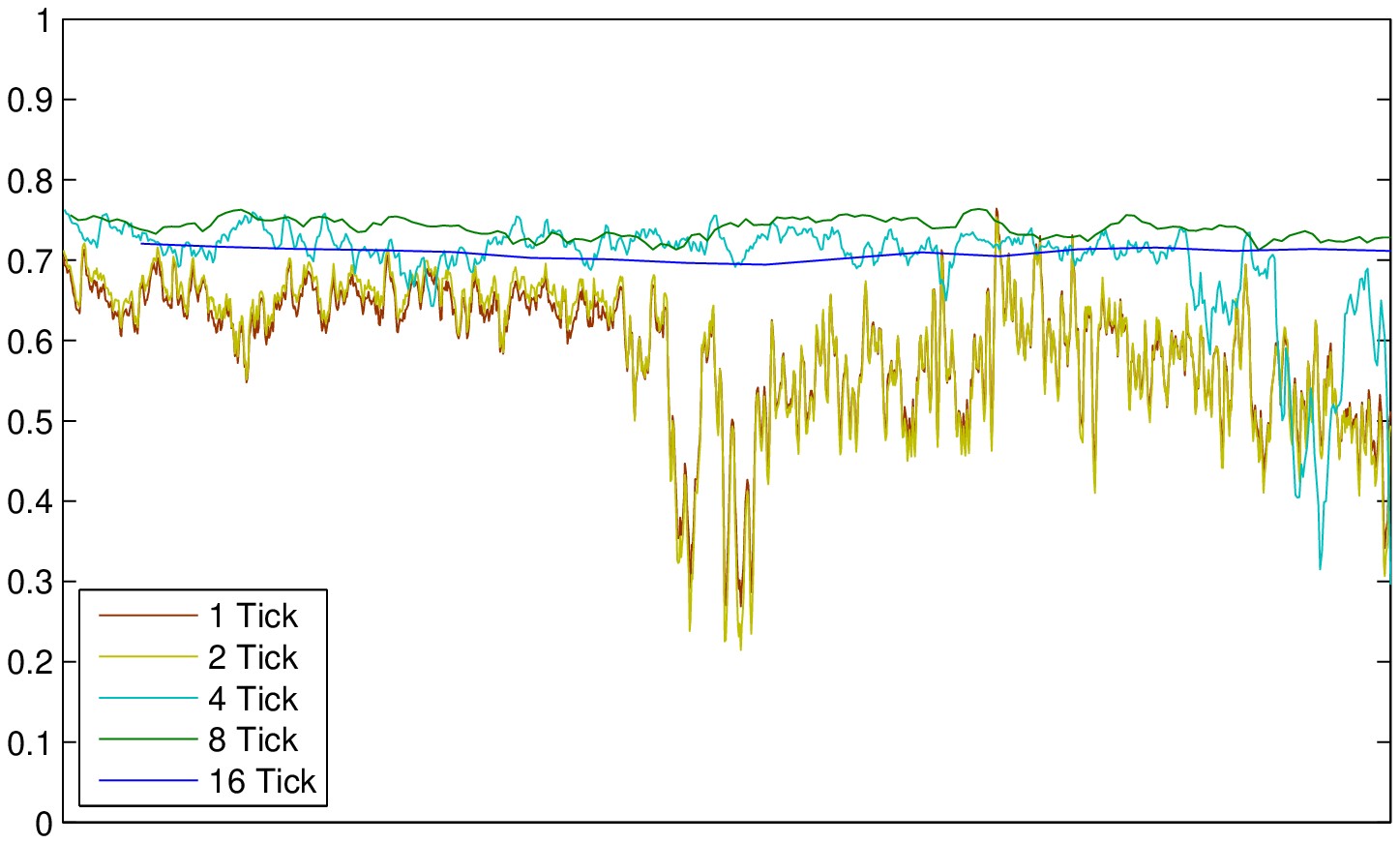}
\put(-270,0){$January2003$}\put(-270,10){$\uparrow$}\put(-30,0){$January2004$}\put(-30,10){$\uparrow$}
\setlength{\unitlength}{1pt}
\caption{\label{fig5}}{Normalized complexity of USD/JPY as a function of time for different sampling rate, e.g. for 16 ticks plot, the first data point is taken and the other 15 data points are discarded.}
\end{center}
\end{figure}

 In an efficient market, the future price is 
determined entirely by the information which is not contained in the price series. 
This means that price changes are random, and 
the variety of patterns are at the maximum level.  Financial time series may behave differently on 
different time scales \cite{BOU,MCC,CIN,JAF,FFA}.
We investigate the variety of patterns in a financial
time series on different time scales.
If we enlarge the sampling interval time from tick to tick to higher 
intervals, the sampling rate would decrease, 
and as a result high frequency contributions in the original time series will be removed. 
By high frequencies we mean those
with time change less than the new sampling interval time.
Finally, we compute the contributions of 
high frequencies in the complexity, for 
$n=7$, See fig. \ref{fig5}. 
It is seen that removing high frequencies leads to increasing 
the complexity in the time period.  This means that in larger scales, 
by the variety
of patterns, these time series behave more similar to a complete random series, 
which is already known. But it is important to notice that in 2003, for 16 ticks period,  
there is still  a large difference between these time series and a 
completely random series. It should be stated that for the time series we have used,
the bid ask bounce effect 
appears only on 1 tick scale.
Therefore, this market behaves inefficiently in both the micro and 
the macro scales; however, in the  micro scale the market is 
more inefficient.

To test the accuracy of the theory of fading dominant patterns, and 
to investigate how the variety of patterns evolve 
in time, data for USD/JPY, EUR/GBP and GBP/CHF for 2003, 2009 and 2012 have been also analyzed. 
Normalized complexity of permutation patterns, $h_7$, for first difference of 
exchange rates are computed and shown in tables 1, 2, and 3.

The sizes of data used in the tables are different. 
Consider the data with minimum size. 
Using Monte Carlo method, we have seen that the parameter 
$h_7$, for a sample realizations of random walk with that minimum size, in 99\% of cases is more than 0.98. So the number 0.98 could be taken as a critical value for rejection/acceptance 
ratio of the {\it Efficient Market
Hypothesis}.

As it is seen from the figures, in 2012 the variety of patterns evolve 
through time , and for 4 ticks scale there are no dominant 
patterns \emph{i.e.}, the scale over
which the target markets have no dominant patterns, 
decreases steadily over time with the emergence of higher frequency
trading.  Although there are still many attempts by the traders to find 
dominant patterns in micro scales; from our
analysis we see that searching for dominant patterns is useless for 
liquid markets.

$$\vbox{
\offinterlineskip
\halign{
\strut \vrule # & \vrule # & \vrule # & \vrule # &\vrule #&\vrule # &\vrule #\vrule\cr
\noalign{\hrule}
 &\, 1 Tick &\, 2 Ticks &\, 4 Ticks&\, 8 Ticks&\, 32 Ticks&\, 128 Ticks   \cr
\noalign{\hrule}
\, USD/JPY &\, 0.668 &\, 0.743 &\, 0.807 &\, 0.862 &\, 0.938 &\, 0.98  \cr
\noalign{\hrule}
\, USD/GBP &\, 0.596 &\, 0.732 &\, 0.783 &\, 0.843 &\, 0.949 &\, 0.98  \cr
\noalign{\hrule}
\, GBP/CHF &\, 0.88 &\, 0.992 &\, 0.954 &\, 0.98 &\, 0.99 &\, 0.99  \cr
\noalign{\hrule}
}
}$$
\vskip -1cm
$$
{\rm Table}\,  1:\, {\rm Normalized}\,  {\rm Complexity}\,  {\rm of}\,  {\rm permutation}\,  {\rm patterns}\, (2006)
$$

$$\vbox{
\offinterlineskip
\halign{
\strut \vrule # & \vrule # & \vrule # & \vrule # &\vrule #&\vrule # \vrule \cr
\noalign{\hrule}
 &\, 1 Tick &\, 2 Ticks &\, 4 Ticks&\, 8 Ticks&\, 32 Ticks \cr
\noalign{\hrule}
\, USD/JPY &\, 0.676 &\, 0.860 &\, 0.922 &\, 0.965 &\, 0.99 \cr
\noalign{\hrule}
\, USD/GBP &\, 0.705 &\, 0.821 &\, 0.885 &\, 0.933 &\, 0.98 \cr
\noalign{\hrule}
\, GBP/CHF &\, 0.795 &\, 0.885 &\, 0.930 &\, 0.960 &\, 0.99 \cr
\noalign{\hrule}
}
}$$
\vskip -1cm
$$
{\rm Table}\,  2:\, {\rm Normalized}\,  {\rm Complexity}\,  {\rm of}\,  {\rm permutation}\,  {\rm patterns}\,  (2009)
$$

$$\vbox{
\offinterlineskip
\halign{
\strut \vrule # & \vrule # & \vrule # & \vrule # \vrule \cr
\noalign{\hrule}
 &\, 1 Tick &\, 2 Ticks &\, 4 Ticks \cr
\noalign{\hrule}
\, USD/JPY &\, 0.924 &\, 0.98 &\, 0.99  \cr
\noalign{\hrule}
\, USD/GBP &\, 0.914 &\, 0.970 &\, 0.99 \cr
\noalign{\hrule}
\, GBP/CHF &\, 0.982 &\, 0.99 &\, 0.99  \cr
\noalign{\hrule}
}
}$$
\vskip -1cm
$$
{\rm Table}\,  3:\, {\rm Normalized}\,  {\rm Complexity}\,  {\rm of}\,  {\rm permutation}\,  {\rm patterns}\,(2012)
$$

\section{Conclusion}
Variety of micro patterns in a financial time series is an important 
measure, and investigating it may provide useful information about the target market. 
In this article it is shown that, permutation entropy may be taken as a useful 
criteria for measuring variety of micro patterns which appears in financial time series.
The permutation 
entropies of order $n$ for some financial 
time series are linear functions of $n$.  In these cases, 
the definition of permutation entropy per symbol, $h_n$, is a useful parameter. 
As $h_n$ has a rapid convergence in terms of $n$, one may use 
a finite $h_n$
 instead of $h_\infty$; we have used $h_7$. 
This makes the method to be very fast. 
The dynamics of the parameter $h_n$  may represent the dynamics of complexity of 
financial time series over time. 
It could also be used as a measure to compare 
complexities of financial time series. 
The complexity of a financial time series may be compared with that of 
a completely random series.  We believe that in recent years, 
investigations of micro patterns and searching for dominant patterns by 
participants in the markets, have faded those patterns. 
As an application, this method is used for three exchange rates.
It is seen that the variety of micro patterns have
evolved between 2006 and 2012; and the scale over which the target markets 
have no dominant patterns, have decreased steadily over time with 
the emergence of higher frequency trading.
Although there are still many attempts for finding dominant patterns in micro scales by 
traders, it seems that searching for dominant patterns 
is useless for liquid markets.


\vskip1cm
\textbf{Acknowledgement}

\noindent  We would like to thank M.~Khorrami, M.~R.~Rahimi~Tabar and F.~Ashtiani for fruitful discussions. We also thank A. Shariati for careful reading of the manuscript.

\end{document}